\documentclass[conference]{IEEEtran}
\IEEEoverridecommandlockouts
\usepackage[numbers,comma,sort&compress]{natbib}
\usepackage{amsmath,amssymb,amsfonts}
\usepackage{algorithmic,soul}
\usepackage{graphicx}
\usepackage{textcomp}
\usepackage{xcolor}

\usepackage{hyperref}

\usepackage{booktabs}
\usepackage{gensymb}
	\usepackage{multirow}
	\usepackage{sidecap}
	\usepackage{graphicx}

\def\BibTeX{{\rm B\kern-.05em{\sc i\kern-.025em b}\kern-.08em
    T\kern-.1667em\lower.7ex\hbox{E}\kern-.125emX}}
\begin{document}

\title{Full-reference Video Quality Assessment for User Generated Content Transcoding\\
\thanks{The authors thank the funding from Tencent (US), University of Bristol, and the UKRI MyWorld Strength in Places Programme (SIPF00006/1).}
}

\author{\IEEEauthorblockN{Zihao Qi\textsuperscript{\dag}, Chen Feng\textsuperscript{\dag}, Duolikun Danier\textsuperscript{\dag},  Fan Zhang\textsuperscript{\dag}, Xiaozhong Xu\textsuperscript{\S}, Shan Liu\textsuperscript{\S} and David Bull\textsuperscript{\dag}}
\IEEEauthorblockA{  \textsuperscript{\dag}\textit{Visual Information Laboratory, University of Bristol, Bristol, UK, BS1 5DD} \\
  \{zihao.qi, chen.feng, duolikun.danier, fan.zhang, dave.bull\}@bristol.ac.uk  \\
  \textsuperscript{\S}\textit{Tencent Media Lab, Palo Alto, CA 94306, USA} \\
  \{xiaozhongxu, shanl\}@tencent.com}

}

\maketitle

\begin{abstract} 

Unlike video coding for professional content, the delivery pipeline of User Generated Content (UGC) involves transcoding where unpristine reference content needs to be compressed repeatedly.
In this work, we observe that existing full-/no-reference quality metrics fail to accurately predict the perceptual quality difference between transcoded UGC content and the corresponding unpristine references. Therefore, they are unsuited for guiding the rate-distortion optimisation process in the transcoding process. 
In this context, we propose a bespoke full-reference deep video quality metric for UGC transcoding. The proposed method features a transcoding-specific weakly supervised training strategy employing a quality ranking-based Siamese structure. The proposed method is evaluated on the YouTube-UGC VP9 subset and the LIVE-Wild database, demonstrating state-of-the-art performance compared to existing VQA methods. The source code of the developed quality metric and the associated training data are available from \url{https://zihaoq1.github.io/FRUGC/}.
\end{abstract}


\section{Introduction}\label{sec:intro}

\begin{figure}[t]
  \centering
  \includegraphics[width=\linewidth]{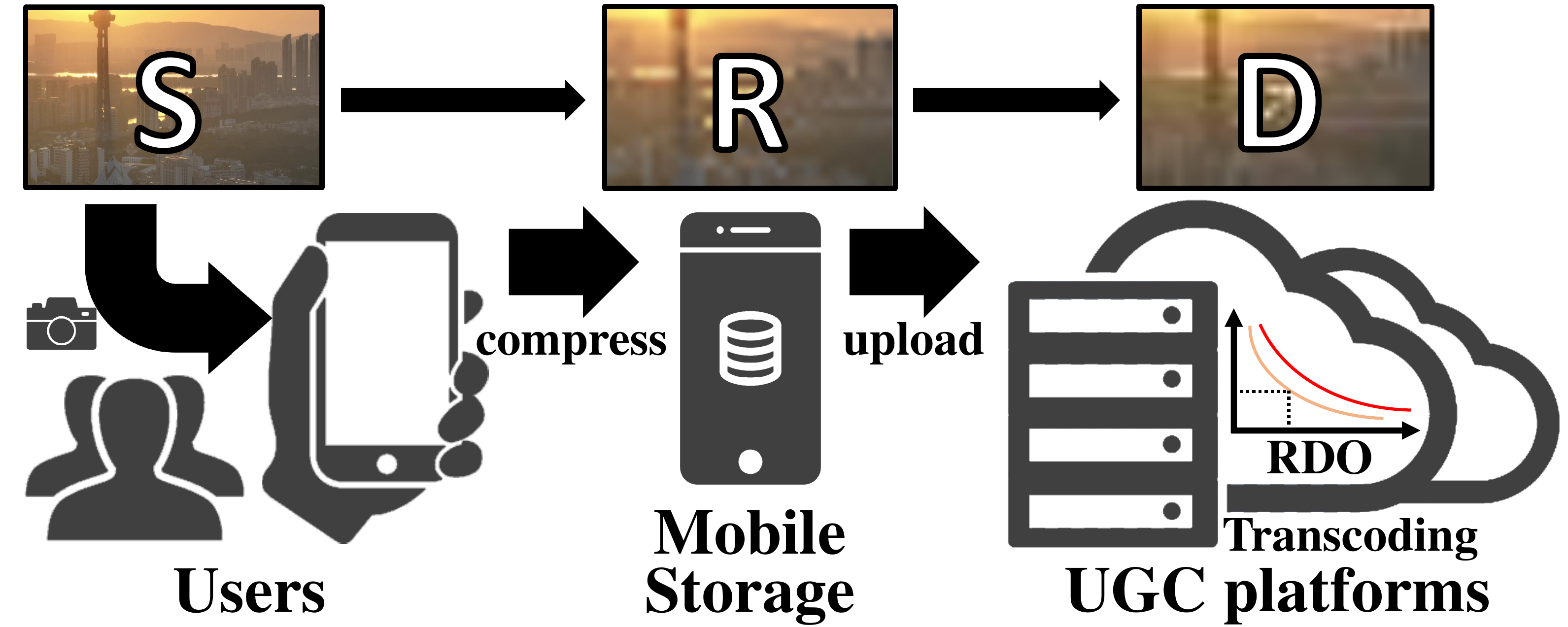}
  \caption{Illustration of the UGC video delivery pipeline. The source content ($\mathbf{S}$) captured by a user is directly compressed on the user device for storage. These videos, denoted by distorted references ($\mathbf{R}$), are then uploaded onto UGC streaming platforms and further compressed into transcoded videos ($\mathbf{D}$) before being transmitted to the viewer.}
\label{fig:transcoding}
\vspace{-10pt}
\end{figure}

Recent advances in video technology alongside rapidly increasing user numbers on various social media platforms have resulted in an explosion of User Generated Content (UGC) on the internet. Before being uploaded to video platform servers, many user generated source videos will exhibit distortions caused by inadequate photography skills, unprofessional equipment, and/or video compression on capture devices~\cite{zhang2023md}. Once uploaded onto a steaming platform, the distorted source content is further transcoded and distributed to viewers, as shown in Fig.~\ref{fig:transcoding}. When the UGC content is transcoded, existing video coding systems perform rate-distortion optimisation (RDO) to achieve maximum quality fidelity for a given bit rate. Crucial to the success of such RDO process is a full reference video quality assessment (VQA) method that can accurately measure the quality degradation of the transcoded version compared to its corresponding unpristine original video.

Existing video quality metrics can be divided into two major categories according to the availability of reference content: full-reference (FR) and no-reference (NR)\footnote{Although reduced reference (RR) VQA methods do exist in the literature, they are not commonly deployed in practical applications, in particular in the context of UGC.}. FR methods aim to capture the quality difference between the reference and distorted videos, which is inherently suitable for guiding the RDO in video coding. However, existing methods~\cite{wang2004image,wang2003multiscale,vu2011spatiotemporal,seshadrinathan2009motion,li2016toward,zhang2018unreasonable,xu2020c3dvqa,feng2022rankdvqa}, either conventional or learning-based, assume pristine reference content, i.e. a lack of distortion in the reference. This assumption is often reflected in model design which relies on natural scenestatistics~\cite{li2016toward,kim2018deep} and in training datasets which only include pristine references. For UGC coding, however, reference videos are often not pristine and can exhibit significant levels of distortion. Hence the use of conventional FR metrics may not be appropriate and could potentially lead to suboptimum performance (as shown in TABLE~\ref{tab1}).

\begin{figure*}[t]
\footnotesize
\begin{minipage}[b]{0.6\linewidth}
  \centering
  \centerline{\includegraphics[width=\textwidth]{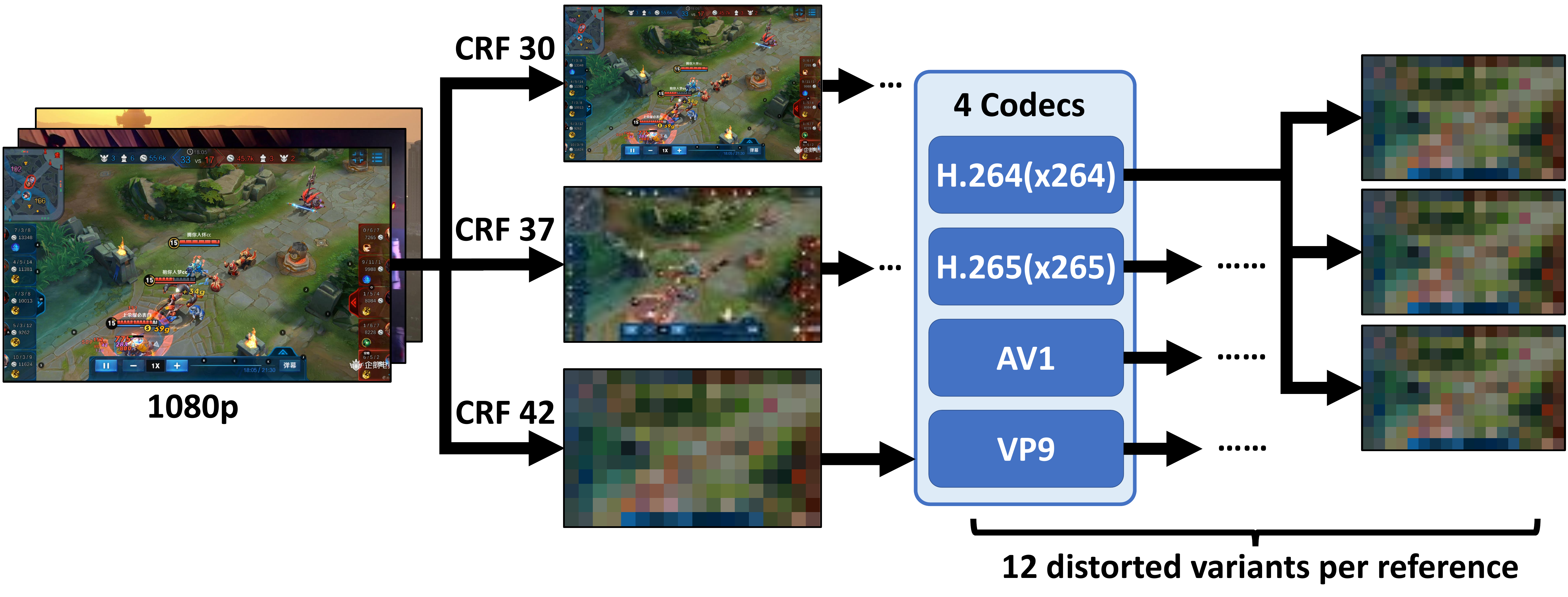}}
(A)
\label{fig3}
\end{minipage}
\begin{minipage}[b]{0.37\linewidth}
  \centering
  \centerline{\includegraphics[width=\textwidth]{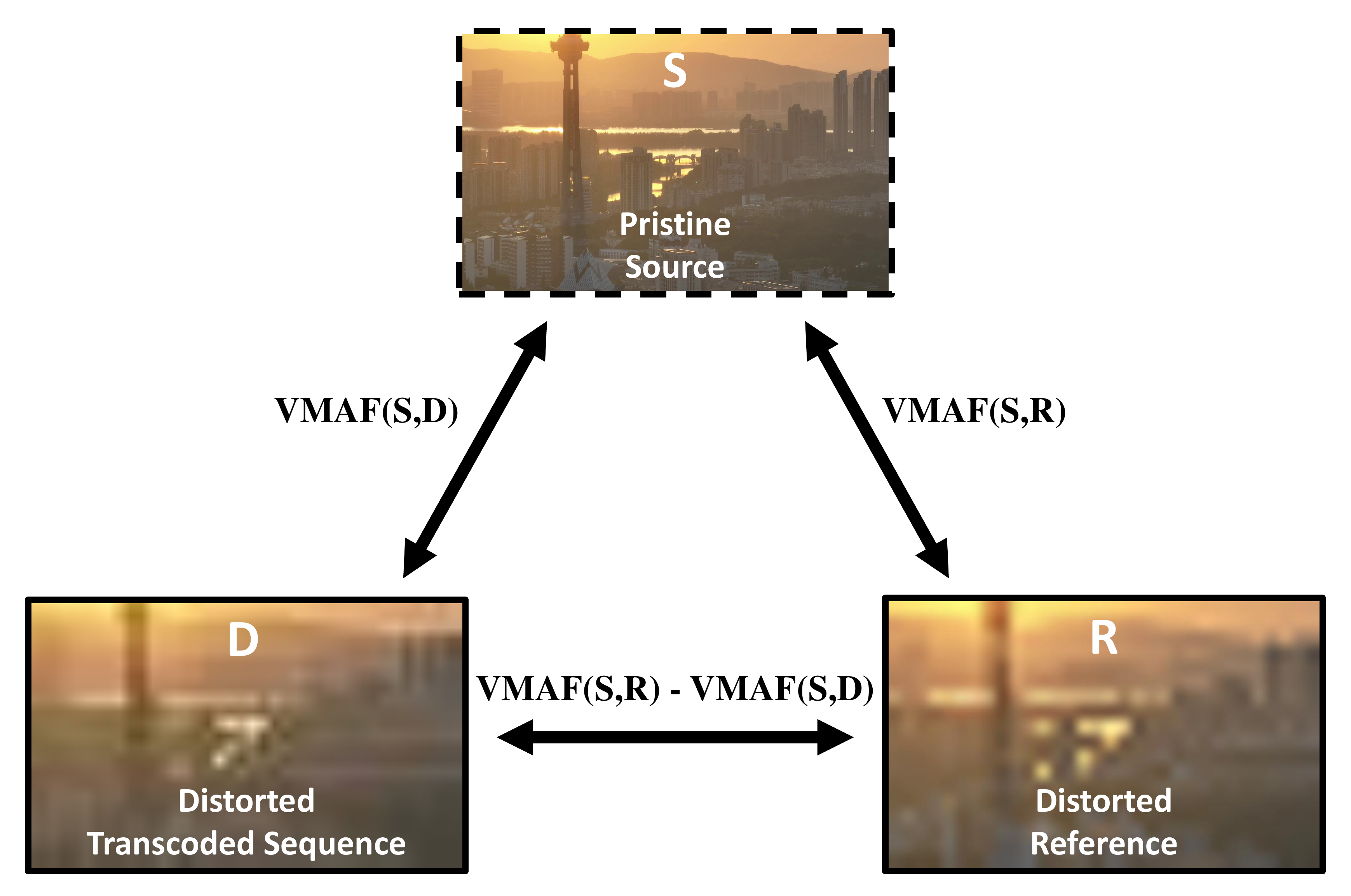}}
(B)
\label{fig2}
\end{minipage}
\caption{(A) The generation of the training database for optimising the PQANet.
(B) The illustration on how VMAF was used in the quality label annotation.}
\label{fig:trinity}
\vspace{-10pt}
\end{figure*}

NR VQA methods~\cite{mittal2012making,saad2014blind,mittal2012no,tu2021ugc,sun2022deep}, on the other hand, aim to provide an absolute quality index for the distorted videos without considering the reference. 
In this case the artefacts inherited from the unpristine reference are considered alongside those arising from transcoding. However,  due to the absence of a reference, the precision of NR methods can be inherently low in terms of measuring the perceptual degradation relative to the reference sequence. Additionally, taking the NR quality index difference between the unpristine reference and the reconstructed sequence can also be ineffective  (as shown in TABLE~\ref{tab1}) due to the accumulation of quality estimation errors.

In this paper we investigate the unique characteristics of UGC full-reference VQA methods for transcoding applications, and expose the need for further research through quantitative experiments. To address this, we propose a FR VQA model, which is trained in a weakly-supervised learning manner based on a Siamese structure. To enable this training methodology, we have developed a new large-scale database with associated quality labels annotated using a proxy quality metric, VMAF. The resulting model was evaluated through a quantitative experiment and compared with both conventional and state-of-the-art FR/NR metrics. Our results confirm the unsatisfactory performance of existing VQA metrics in the targeted scenario of UGC transcoding, and demonstrate the evident improvement achieved by the proposed method. 


\section{Proposed Algorithm}\label{sec:pro}

This section first describes the unique characteristics of full reference video quality assessment for UGC transcoding, then presents a new training strategy for optimising a deep VQA network which predicts the quality difference between a transcoded video and its corresponding unpristine reference. This training strategy has inspired the development of a new large training database with reliable quality annotations for FR VQA in the context of UGC transcoding.

\subsection{The UGC Transcoding Pipeline}
\label{sec:pipeline}

In the UGC delivery pipeline of Fig. \ref{fig:transcoding}, the video is processed in three stages. When a video is captured, typically using a mobile device, the ``pristine source'', $\mathbf{S}$, is not stored losslessly. Instead, it is directly compressed by a video codec integrated within the device, e.g, by H.264/AVC~\cite{h264} or H.265/HEVC~\cite{h265}, to generate a version with much lower bit rate, denoted as the ``distorted reference'', $\mathbf{R}$. This signal tends to contain visible compression artefacts compared to the original uncompressed UGC video. These distorted references are then uploaded by users to a UGC video platform, where they are further transcoded into distorted versions, denoted as the ``distorted transcoded sequence'', $\mathbf{D}$ (another layer of compression), before distribution to viewers across variable bandwidth networks. While conventional FR VQA focuses on the single-layer compression scenario, where the pristine $\mathbf{S}$ is encoded into $\mathbf{R}$, here we target the transcoding  from $\mathbf{R}$ to $\mathbf{D}$.

\subsection{The Employed Network Architecture}

We use the network architecture proposed in \cite{feng2022rankdvqa}. As illustrated by Fig. \ref{fig:training}. (A), this network first takes a pair of transcoded and reference patches, $\mathbf{P}_{\mathbf{D}_1}$ and $\mathbf{P}_{\mathbf{R}_1}$, as input, which correspond to the co-located 256$\times$256$\times$12 (H$\times$W$\times$T) video patches extracted from the transcoded video $\mathbf{D}_1$ and its distorted reference $\mathbf{R}_1$ respectively. The input patches are then processed in two stages. In Stage 1, a transformer-based patch-wise quality assessment network (PQANet) is employed to output a quality index, $Q_{(\mathbf{P}_{\mathbf{R}_1}, \mathbf{P}_{\mathbf{D}_1})}$. In the second stage, the quality indices generated by PQANet for all the video patches from the same transcoded and reference videos are then passed to the Spatio-Temporal Aggregation Network (STANet) to obtain the sequence level quality index for $\mathbf{D}_1$ (against $\mathbf{R}_1$). Details on network architecture designs can be found in \cite{feng2022rankdvqa}.

\subsection{The Training Strategy}

It is noted that \cite{feng2022rankdvqa} uses a ranking-inspired training methodology which allows the use of a proxy quality metric (VMAF \cite{li2016toward} in this case) to generate reliable quality labels for the training content. This enables the creation of a large and diverse database to improve the generalisation of the deep VQA models without performing expensive and extensive subjective experiments. However, this training methodology cannot be directly applied here, because existing quality metrics such as VMAF cannot provide a robust quality prediction if the reference content contains various UGC artefacts  (this has been confirmed by \cite{wang2019youtube} and by our results in TABLE \ref{tab1}).

\begin{figure*}[t]
\footnotesize
\centering
\begin{minipage}[b]{0.58\linewidth}
  \centering
  \centerline{\includegraphics[width=\textwidth]{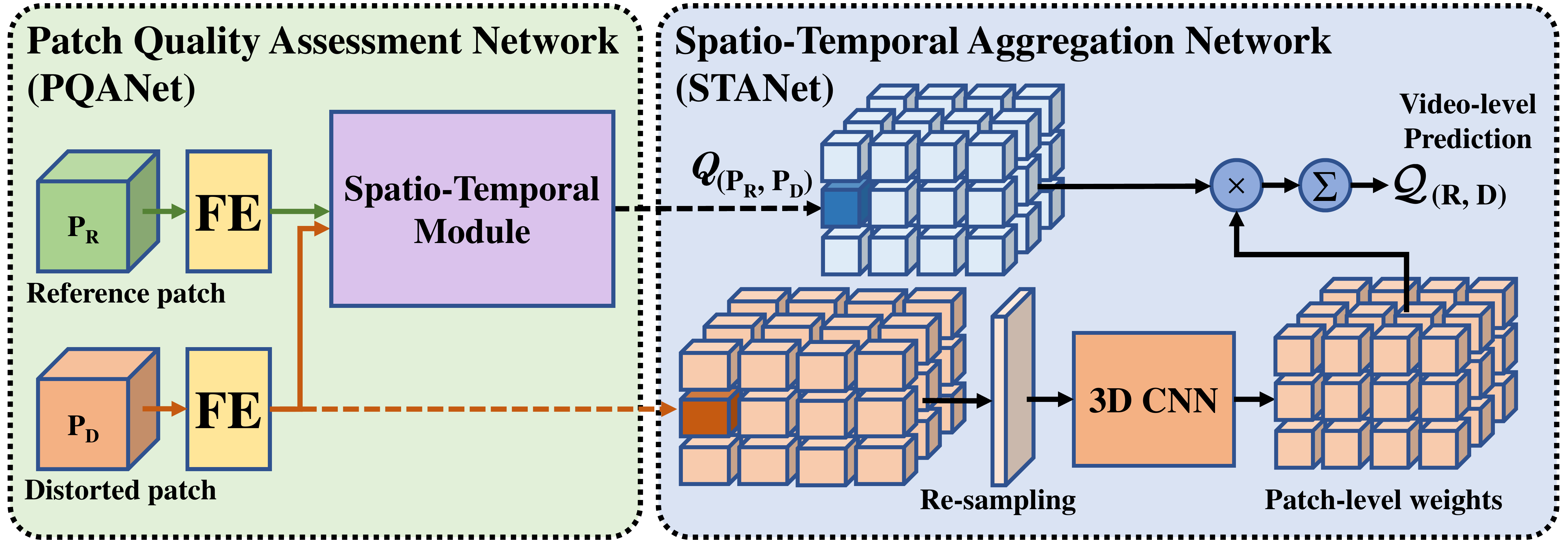}} 
(A)
\end{minipage}
\begin{minipage}[b]{0.32\linewidth}
  \centering
  \centerline{\includegraphics[width=\textwidth]{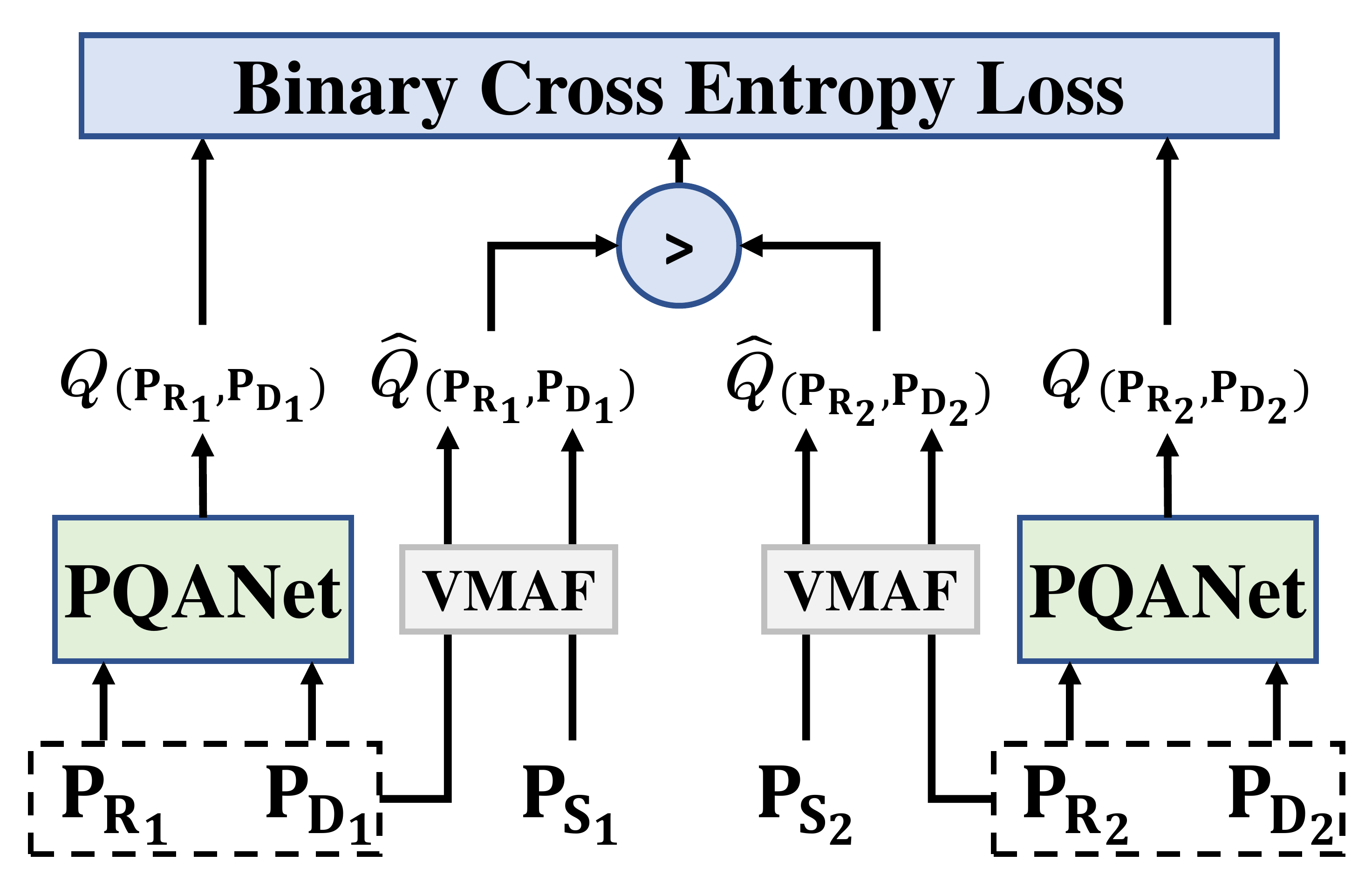}}
(B)
\end{minipage}

\caption{(A) The overall picture of the inference workflow. (B) The illustration of the proposed training strategy.}
\label{fig:training}
\vspace{-10pt}
\end{figure*}

To address this issue in our training process, we additionally employ the pristine source, $\mathbf{P}_{\mathbf{S_1}}$ and $\mathbf{P}_{\mathbf{S_2}}$, for quality labelling to improve the annotation reliability, as shown in Fig. \ref{fig:training}. (B). It is noted that these pristine source versions are not input to the network in either training or inference stages.

Specifically, during the training process, based on the ranking-inspired training method used in \cite{hou2022perceptual,feng2022rankdvqa}, two pairs of video patches, $(\mathbf{P}_{\mathbf{R}_1},\mathbf{P}_{\mathbf{D}_1})$ and $(\mathbf{P}_{\mathbf{R}_2},\mathbf{P}_{\mathbf{D}_2})$ are input into PQANet and, based on their outputs $Q_{(\mathbf{P}_{\mathbf{R}_1},\mathbf{P}_{\mathbf{D}_1})}$ and $ Q_{(\mathbf{P}_{\mathbf{R}_2},\mathbf{P}_{\mathbf{D}_2})}$, the probability $p$ of patch $\mathbf{P}_{\mathbf{D}_1}$ being of higher quality (with respect to its reference counterpart) than $\mathbf{P}_{\mathbf{D}_2}$ is obtained using a sigmoid function:
\begin{equation}
    p = \mathrm{sigmoid}(Q_{(\mathbf{P}_{\mathbf{R}_1},\mathbf{P}_{\mathbf{D}_1})} - Q_{(\mathbf{P}_{\mathbf{R}_2},\mathbf{P}_{\mathbf{D}_2})}).
\end{equation}

To obtain the training targets, we also employ VMAF \cite{li2016toward} as a proxy metric here, but based on a different approach compared to \cite{feng2022rankdvqa}. Specifically, we calculate the VMAF values, $\mathrm{VMAF}(\mathbf{P}_{\mathbf{S}_1},\mathbf{P}_{\mathbf{D}_1})$, $\mathrm{VMAF}(\mathbf{P}_{\mathbf{S}_2},\mathbf{P}_{\mathbf{D}_2})$, $\mathrm{VMAF}(\mathbf{P}_{\mathbf{S}_1},\mathbf{P}_{\mathbf{R}_1})$ and $\mathrm{VMAF}(\mathbf{P}_{\mathbf{S}_2},\mathbf{P}_{\mathbf{R}_2})$ separately. Based on these, we then obtain the quality differences, $ \hat{Q}_{(\mathbf{P}_{\mathbf{R}_1},\mathbf{P}_{\mathbf{D}_1})}$ and $ \hat{Q}_{(\mathbf{P}_{\mathbf{R}_2},\mathbf{P}_{\mathbf{D}_2})}$ between the unpristine reference and the distorted content for both patch pairs:
    \begin{equation}
    \left\{
    \begin{array}{l}
          \hat{Q}_{(\mathbf{P}_{\mathbf{R}_1},\mathbf{P}_{\mathbf{D}_1})} = \mathrm{VMAF}(\mathbf{P}_{\mathbf{S}_1},\mathbf{P}_{\mathbf{R}_1}) - \mathrm{VMAF}(\mathbf{P}_{\mathbf{S}_1},\mathbf{P}_{\mathbf{D}_1}) \\
          \hat{Q}_{(\mathbf{P}_{\mathbf{R}_2},\mathbf{P}_{\mathbf{D}_2})} = \mathrm{VMAF}(\mathbf{P}_{\mathbf{S}_2},\mathbf{P}_{\mathbf{R}_2}) - \mathrm{VMAF}(\mathbf{P}_{\mathbf{S}_2},\mathbf{P}_{\mathbf{D}_2})  \\
    \end{array}
    \right.
    \label{eq:deltaQ}
\end{equation}
It is noted that in the VMAF calculations above, as we always use pristine content as references, the accuracy of the quality prediction has been maintained. This process has also been illustrated by Fig. \ref{fig:trinity}.(B).

As in \cite{feng2022rankdvqa}, rather than minimising the difference between the quality difference $ \hat{Q}$ and the network output $Q$, we further obtain the quality ranking information $r$ between two quality difference values $ \hat{Q}_{(\mathbf{P}_{\mathbf{R}_1},\mathbf{P}_{\mathbf{D}_1})}$ and $ \hat{Q}_{(\mathbf{P}_{\mathbf{R}_2},\mathbf{P}_{\mathbf{D}_2})}$:
\begin{equation}
    r = \left\{
    \begin{array}{l}
         1, \text{if }   \hat{Q}_{(\mathbf{P}_{\mathbf{R}_1},\mathbf{P}_{\mathbf{D}_1})}- \hat{Q}_{(\mathbf{P}_{\mathbf{R}_2},\mathbf{P}_{\mathbf{D}_2})} >\sigma \\
         0, \text{if }   \hat{Q}_{(\mathbf{P}_{\mathbf{R}_1},\mathbf{P}_{\mathbf{D}_1})}- \hat{Q}_{(\mathbf{P}_{\mathbf{R}_2},\mathbf{P}_{\mathbf{D}_2})} <-\sigma\\ 
    \end{array}
    \right.
    \label{eq:r}
\end{equation}
Here $\sigma$ is a threshold, which is based on the ranking ability of $\hat{Q}_{(\cdot,\cdot)}$. For cases when $-\sigma\leq r \leq \sigma$, we have excluded them in the database generation process. More detailed definition of $\sigma$ and its values are described in Section \ref{sec:datageneration}.  Based on this, we can train the network with the binary cross entropy loss:
\begin{equation}
    \mathcal{L} = -(r\cdot\mathrm{log}p + (1-r)\cdot\mathrm{log}(1-p)).
\end{equation}
This training strategy is illustrated by Fig. \ref{fig:training}.(B).

The training methodology for STANet remains the same as that in the original RankDVQA paper~\cite{feng2022rankdvqa}. The only change is the training database, as described below.

\subsection{Training Databases}\label{sec:datageneration}

\textbf{Training database for Stage 1.} For the training of PQANet, to closely simulate the UGC transcoding pipeline described in Section \ref{sec:pipeline}, we firstly collated 252 pristine source sequences from the BVI-DVC dataset~\cite{ma2021bvi}, the CVPR 2022 CLIC challenge training dataset~\cite{clic2022} and the YouTube-UGC 2K database~\cite{wang2019youtube} (non-overlapping with VP9 subset) alongside 6 self-captured videos. These 258 original sequences ($\mathbf{S}$) were then compressed with x264~\cite{ffmpeg} (CRF = 30, 37, 42, medium preset), a commonly used video codec on mobile devices, to generate distorted reference $\mathbf{R}$. This is to simulate the compression process on user-capture devices. Each distorted reference video was further compressed into 12 transcoded sequences ($\mathbf{D}$) at three quantisation levels, by 4 different codecs (x264, x265~\cite{ffmpeg}, AV1~\cite{aom}, VP9~\cite{vp9}).  This emulates the transcoding process on UGC platforms. The workflow is illustrated in Fig.~\ref{fig:trinity} (A). This results in 9,288 distorted sequences.

Based on the generated transcoded sequences described above and their associated distorted references ($\mathbf{R}$), we adopted the same patch generation method as in \cite{feng2022rankdvqa}, which randomly crops 256$\times$256$\times$12 spatio-temporal patches. We further combined two different patch groups as a training instance $\{(\mathbf{P}_{\mathbf{R}_1},\mathbf{P}_{\mathbf{D}_1}), (\mathbf{P}_{\mathbf{R}_2},\mathbf{P}_{\mathbf{D}_2}),r\}$ to enable ranking-based optimisation. In each training instance, two ($\mathbf{P}_\mathbf{R},\mathbf{P}_\mathbf{D}$) pairs can either correspond to the same distorted reference sequence, denoted by single source (SS), or to a different distorted reference sequence, labelled as dual source (DS). The label $r$ is obtained by Equation (\ref{eq:deltaQ}) and (\ref{eq:r}).

\begin{figure}[t]
\hfill
\centering
\begin{minipage}{0.49\linewidth}
  \centering
  \centerline{\includegraphics[width=\textwidth]{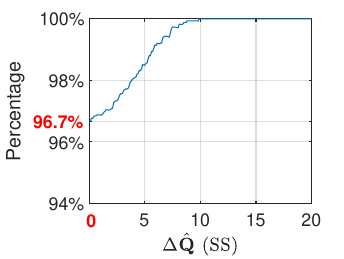}}
\end{minipage}
\begin{minipage}{0.49\linewidth}
  \centering
  \centerline{\includegraphics[width=\textwidth]{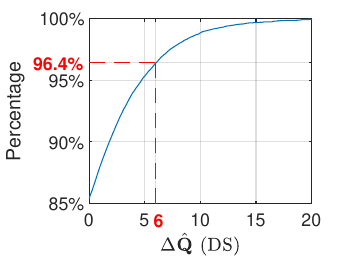}}
\end{minipage}
\caption{$\hat{\mathcal{Q}}$ difference (see Equation (\ref{eq:r})) stands for the difference between two $\hat{\mathcal{Q}}$ values (on video-level).  Percentage: the accuracy ratio of $\hat{\mathcal{Q}}$ difference-based ranking results that are consistent with human judgement. Results produced on ICME database~\cite{wang2021challenge}. (Left) Single sources  (SS). (Right) Dual sources (DS).}
\label{fig:vmaf}
\vspace{-10pt}
\end{figure}

To ensure the reliability of the VMAF-based training labels here, we followed the approach described in \cite{hou2022perceptual,feng2022rankdvqa} to evaluate the ranking ability of $\hat{Q}_{(\cdot,\cdot)}$ in Equation (\ref{eq:r}) and determine the value of the threshold $\sigma$, in the context of the UGC transcoding scenario based on the ICME Challenge database \cite{wang2021challenge}. It contains 900 unpristine references alongside 6,300 transcoded sequences (excluding the testing set). Specifically, we calculate the accuracy of $\hat{Q}_{(\cdot,\cdot)}$ when it is used to differentiate quality differences between every two pairs of transcoded sequences and their unpristine references (the same or different) based on the subjective ground truth. The results are shown in Fig. \ref{fig:vmaf}. It can be observed that, when the 
$\hat{Q}_{(\cdot,\cdot)}$ predicted quality difference, $|\hat{Q}_{(\mathbf{P}_{\mathbf{R}_i},\mathbf{P}_{\mathbf{D}_i})}- \hat{Q}_{(\mathbf{P}_{\mathbf{R}_j},\mathbf{P}_{\mathbf{D}_j})}|$ (following Equation (\ref{eq:deltaQ})), is larger than 0 (for the single source scenario) or 6 (for the dual sources), the ranking accuracy according to the actual subjective score is above 96\% (which provides a good trade off between the number of training instances and reliability). These two thresholds are hence used to examine the labelled patches - for each training instance; if the absolute difference in $\hat{Q}_{(\cdot,\cdot)}$ is smaller than these thresholds, we exclude it. Otherwise, we obtain the label according to Equation~(\ref{eq:r}). This results in 315,059 training instances (80\% SS, 20\% DS).

\textbf{Training database for Stage 2. } For STANet, we followed the same training strategy as in \cite{feng2022rankdvqa}, but used the ICME Challenge 2021~\cite{wang2021challenge} database to optimise the spatio-temporal pooling network. Here we expect STANet to take all the patch-level quality indices and output the final sequence-level quality score. Detailed training procedures can be found in \cite{feng2022rankdvqa}.

\subsection{Implementation Details}

Pytorch 1.10 was used to implement both  PQANet and STANet. The training parameters used were: Adam optimisation with hyper-parameters of $\beta1=0.9$ and $\beta2=0.999$; 60 training epochs; batch size of 4; the initial learning rate is 0.0001 with a weight decay of 0.1 after every 20 epochs. Both training and evaluation were executed on a computer with a 2.4GHz Intel CPU and an NVIDIA P100 graphic card.

 \section{Results and Discussion}
 \label{sec:results}


Most existing UGC databases, such as YouTube-UGC~\cite{wang2019youtube}, KoNVid-1k~\cite{hosu2017konstanz}, and LIVE-VQC~\cite{sinno2018large}, are designed for no-reference video quality assessment, and hence cannot be employed here for evaluating full reference quality metrics. Therefore we conducted a benchmark experiment based on two UGC video quality databases with unpristine references, YouTube-UGC VP9 subset~\cite{wang2019youtube} and LIVE-WILD~\cite{yu2021predicting}, which can simulate the UGC transcoding scenario.  The YouTube-UGC VP9 subset is part of YouTube-UGC~\cite{wang2019youtube} database, containing 507 transcoded sequences compressed by VP9 based on 169 reference sequences with visual artefacts. LIVE-WILD consists of 220 distorted sequences which are derived from 55 unpristine reference sequences. It is noted that the overall perceptual quality of the reference content in the LIVE WILD database is higher than that of the YouTube-UGC VP9 subset. This implies that the latter may be more challenging when used for evaluating FR VQA methods.


Our proposed method was tested against 7 existing FR VQA methods (PSNR, SSIM, MS-SSIM~\cite{wang2003multiscale}, LPIPS~\cite{zhang2018unreasonable}, VMAF~\cite{li2016toward}, C3DVQA~\cite{xu2020c3dvqa} and RankDVQA~\cite{feng2022rankdvqa}), and 6 NR quality metrics (NIQE~\cite{mittal2012making}, BRISQUE~\cite{mittal2012no}, VIIDEO~\cite{mittal2015completely}, VBLIINDS~\cite{saad2014blind}, VIDEVAL~\cite{tu2021ugc} and SimpleVQA~\cite{sun2022deep}). Here PSNR, SSIM, MS-SSIM, NIQE, BRISQUE, VIIDEO and VBLIINDS  are conventional VQA methods, while VMAF is a regression-based quality metric. C3DVQA, RankDVQA, VIDEVAL and SimpleVQA are deep learning-based VQA methods, and their pre-trained models were used for benchmarking. For all learning-based methods including ours, we did not perform intra database cross-validation, and there is no overlap between the training material and all the sequences in these two benchmark databases.

All FR quality metrics tested are used to predict the quality differences between transcoded videos and their unpristine references. The Spearman Ranking Order Correlation Coefficients (SROCC), and Kendall Ranking Correlation Coefficients (KRCC) between their predicted quality difference scores and the true subjective different mean opinion scores (DMOS) on each database are employed to measure their performance. For each NR metric, we adapt to the FR transcoding scenario to calculate the quality index of a transcoded video and its unpristine reference separately, and then obtain their quality differences to correlate with the DMOS\footnote{We did not evaluate the quality predictions of NR methods on the distorted videos against their MOS values (their corresponding SROCC values are relatively low anyway - up to 0.3882 on YT-UGC VP9 and 0.7113 on LIVE-WILD), because our focus is on the quality differences between transcoded content and unpristine references.}.

\begin{table}[t]
 \caption{FR and NR results on available FR UGC datasets. In each cell, the values x(y) corresponds to the SROCC or KRCC value (x) and the F-test result (y) at 95\% confidence interval. Here y$=1$ indicates that the metric is superior to our proposed method (y$=-1$ if the opposite is true), and y$=0$ means that there is no significant performance difference between them. }
\centering
\resizebox{\linewidth}{!}{
\begin{tabular}{r|l|l|l|l}
\toprule
\multirow{2}{*}{\textbf{Metrics}}& \multicolumn{2}{c|}{\textbf{YT-UGC VP9}} &  \multicolumn{2}{c}{\textbf{LIVE-WILD}} \\
\cmidrule{2-5} 
& \textbf{SROCC} & \textbf{KRCC} & \textbf{SROCC} & \textbf{KRCC} \\
\midrule \multicolumn{1}{l}{\textbf{Full-reference }} \\
\midrule
PSNR & 0.3946 (-1) & 0.2737 (-1)  & 0.7613 (-1) & 0.5687 (-1) \\
SSIM & \underline{0.5384 (-1)} & \underline{0.3752 (-1)}  & 0.8711 (0) & 0.6755 (0) \\
MS-SSIM~\cite{wang2003multiscale} & 0.5290 (-1) & 0.3677 (-1)  & 0.8753 (0) & 0.6788 (0) \\
LPIPS~\cite{zhang2018unreasonable} & 0.4491 (-1) & 0.3087 (-1)  & 0.8587 (0) & 0.6618 (0) \\
VMAF 0.6.1~\cite{li2016toward} & 0.4378 (-1) & 0.3025 (-1)  & \underline{0.8957 (0)} & \underline{0.7127 (0)} \\
C3DVQA~\cite{xu2020c3dvqa} & 0.4732 (-1) & 0.3220 (-1) & 0.6980 (-1) & 0.5157 (-1) \\
RankDVQA~\cite{feng2022rankdvqa} & 0.4725 (-1) & 0.3272 (-1) & 0.8803 (0) & 0.6958 (0) \\
\midrule \multicolumn{1}{l}{\textbf{No-reference }} \\
\midrule
NIQE~\cite{mittal2012making} & 0.1718 (-1) & 0.1136 (-1) & 0.3443 (-1) & 0.2352 (-1) \\ 
BRISQUE~\cite{mittal2012no} & 0.0934 (-1) & 0.0632 (-1) & 0.7297 (-1) & 0.5371 (-1) \\
VIIDEO~\cite{mittal2015completely} & 0.2121 (-1) & 0.1425 (-1) & 0.0201 (-1) & 0.0148 (-1) \\
VBLIINDS~\cite{saad2014blind} & 0.2042 (-1)  & 0.1365 (-1) & 0.3660 (-1) & 0.2529 (-1) \\ 
VIDEVAL~\cite{tu2021ugc} & 0.1868 (-1) & 0.1089 (-1) & 0.3402 (-1) & 0.2291 (-1) \\
SimpleVQA~\cite{sun2022deep} & 0.4741 (-1) & 0.3318 (-1) & 0.8226 (-1) & 0.6277 (-1) \\
\midrule
\midrule
\textbf{Ours} & \textbf{0.6974} & \textbf{0.4951}  & \textbf{0.8975} & \textbf{0.7149} \\ 
\bottomrule
\end{tabular}
 }
 \label{tab1}
\vspace{-5pt}
\end{table}


Table~\ref{tab1} summarises the performance results for all the tested full reference and no reference VQA methods on both databases. It can be observed that the proposed FR quality metric outperforms all other tested FR and NR VQA methods on both databases. Furthermore, the improvement of our method over the second performer on the YT-UGC VP9 dataset is much higher than that based on the LIVE-WILD database. This is most likely because the reference sequences in the former contain more visual artefacts compared to the latter. This makes the YT-UGC VP9 dataset much more challenging, as confirmed by the relatively low SROCC values of all the tested quality metrics (below 0.7). It is also noted that on both databases, the proposed method performs well compared to RankDVQA - this verifies the effectiveness of the new training methodology, which is one of our primary contributions.

Table~\ref{tab1} also provides statistical test results to differentiate the performance between the proposed quality metric and each of the benchmark results. Here we employed an F-test following \cite{zhang2015perception}, between the prediction residuals (after a non-linear fitting) of our method and other benchmark approaches. The results show that the proposed method is significantly better than all the other tested quality metrics on the YT-UGC VP9 database. On the LIVE-WILD database, there is no significant difference between SSIM, MS-SSIM, VMAF, RankDVQA, and our approach, but the latter statistically performs better than all the other FR and NR metrics.

\section{Conclusion}
\label{sec:conclusion}

In this paper, we first justified the need for tailored metrics for assessing transcoded UGC videos, because existing VQA methods cannot accurately predict the quality difference between them and their unpristine references. To address this issue, a new FR VQA method is proposed based on a weakly-supervised Siamese training methodology and a large training dataset with reliable quality annotations. Our results demonstrate the improvement of the proposed method over existing FR and NR quality metrics when tested on two full reference UGC transcoding databases. Future work should focus on further improving the correlation performance by enhancing the architecture for the focused application scenario.

\small
\bibliographystyle{IEEEtran}
\bibliography{IEEEabrv,IEEEexample}

\vspace{12pt}

\end{document}